\documentclass[prl,nofootinbib,preprintnumbers,superscriptaddress,twocolumn]{revtex4-1}

\usepackage{amssymb}
\usepackage{amsmath}
\usepackage{graphicx}
\usepackage{verbatim}
\usepackage{amsfonts}
\usepackage{color}

\newcommand{\Nd}[0]{\mathcal N_{\rm{\scriptstyle{NC}}}^{(d)}}
\newcommand{\Nu}[0]{\mathcal N_{\rm{\scriptstyle{NC}}}^{(u)}}
\newcommand{\Nq}[0]{\mathcal N_{\rm{\scriptstyle{NC}}}^{(q)}}
\newcommand{\Nbc}[0]{\mathcal N_{\rm{\scriptstyle{CC}}}}

\allowdisplaybreaks

\begin{document}

\preprint{LMU-ASC 34/15}
\preprint{FLAVOUR(267104)-ERC-99}

\title{Signatures of a nonstandard Higgs from flavor physics}

\author{Oscar Cat\`a}
\affiliation{Ludwig-Maximilians-Universit\"at M\"unchen, 
   Fakult\"at f\"ur Physik,\\
   Arnold Sommerfeld Center for Theoretical Physics, 
   80333 M\"unchen, Germany}
\author{Martin Jung}
\affiliation{
TUM Institute for Advanced Study, Lichtenbergstr. 2a, D-85747 Garching, Germany.}
\affiliation{
Excellence Cluster Universe, Technische Universit\"at M\"unchen, Boltzmannstr. 2, D-85748 Garching,
Germany.}


\begin{abstract}
We examine the constraints coming from incorporating the full Standard Model gauge symmetry into the
effective field theory description of flavor processes, using semileptonic decays as paradigmatic
examples. Depending on the dynamics triggering electroweak symmetry breaking, different
patterns of correlations between the Wilson coefficients arise.
Interestingly, this implies that flavor experiments are capable of shedding light upon the nature of the Higgs boson without actually requiring Higgs final states. Furthermore, the observed correlations can simplify model-independent analyses of these decays.
\end{abstract}

\maketitle


\noindent{\it{1. Introduction}}. The discovery of the Higgs boson at the LHC~\cite{Aad:2012tfa}
and the current knowledge of its properties have once more confirmed 
that the Standard Model (SM) is an excellent low-energy description of
the electroweak interactions. The precise nature of the Higgs boson -- which is linked to the
characteristics of potential new physics (NP) -- is however an issue that remains to be settled; this is one of the main goals of Run-II at the LHC, using the analysis of Higgs
couplings in multi-Higgs production processes. However, recent studies~\cite{Barr:2014,Azatov:2015}
draw rather pessimistic conclusions regarding the capability of the LHC to discriminate,
{\it{e.g.}}, weakly- from strongly-coupled NP scenarios from double Higgs production.

In this paper we argue that flavor processes
have the potential to test the dynamics triggering electroweak symmetry breaking (EWSB), despite being unable to probe Higgs couplings directly. In
that sense, flavor experiments can provide valuable information on the Higgs and should be seen as complementary to the LHC effort to identify
its properties. Turning the argument around, interesting consequences arise for flavor physics when assuming weakly-coupled NP.

For illustration we will consider heavy-flavor semileptonic processes, specifically the flavor-changing neutral current (FCNC) processes $D\to
D'\ell^+\ell^-$ and $U\to U'\ell^+\ell^-$ as well as the charged-current processes $D\to U \ell\nu$ (effectively including $U\to D\ell\nu$), where
$D^{(\prime)}$ and $U^{(\prime)}$ are down-type and up-type quarks, respectively, and
$\ell=e,\mu,\tau$. For definiteness we will focus on $b\to s\ell\ell$, $c\to u \ell \ell$ and $b\to
c\tau\nu$ as representatives of the different groups. At hadronic scales $\Lambda\ll M_W$ the effective field
theory for down-quark semileptonic processes reads~\cite{Buchalla:1995vs}\footnote{In this work we do not discuss the contributions from four-quark
operators, which are of minor importance in these decays.}
\begin{align}
\mathcal L_{\rm eff}^{b\to s\ell\ell} &= \frac{4G_F}{\sqrt{2}}\lambda_{ts}\frac{e^2}{(4\pi)^2}\sum_{i}^{12}C_i^{(d)}\mathcal
O_i^{(d)}\,,
\end{align}
where $\lambda_{ts}=V_{tb}V_{ts}^*$ and the operators are defined as 
\begin{align}\label{opd}
{\cal{O}}_{7}^{(\prime)}&=\frac{m_b}{e}({\bar{s}}\sigma^{\mu\nu}P_{R(L)}
b)F_{\mu\nu}\,,\!\!\!\!\!\!\!\!\!\!&&\nonumber\\
{\cal{O}}_9^{(\prime)}&=({\bar{s}}\gamma_{\mu}P_{L(R)} b)~{\bar{l}}\gamma^{\mu}l\,,
&{\cal{O}}_{10}^{(\prime)}&=({\bar{s}}\gamma_{\mu}P_{L(R)}
b)~{\bar{l}}\gamma^{\mu}\gamma_5l\,,\nonumber\\
{\cal{O}}_S^{(\prime)}&=({\bar{s}}P_{R(L)} b)~{\bar{l}}l\,,
&{\cal{O}}_P^{(\prime)}&=({\bar{s}}P_{R(L)} b)~{\bar{l}}\gamma_5l\,,\nonumber\\
{\cal{O}}_T&=({\bar{s}}\sigma_{\mu\nu}b)~{\bar{l}}\sigma^{\mu\nu}l\,,
&{\cal{O}}_{T5}&=({\bar{s}}\sigma_{\mu\nu}b)~{\bar{l}}\sigma^{\mu\nu}\gamma_5l\,.
\end{align}
For up-quark transitions, one likewise finds
\begin{align}
\mathcal L_{\rm eff}^{c\to u\ell\ell} &= \frac{4G_F}{\sqrt{2}}\lambda_{bu}\frac{e^2}{(4\pi)^2}\sum_{i}^{12}C_i^{(u)}\mathcal
O_i^{(u)}\,,
\end{align}
where $\lambda_{bu}=V_{cb}V_{ub}^*$ and the operators are obtained by the trivial flavor replacements $(b;s)\to (c;u)$ in Eq.~(\ref{opd}). Finally,
charged-current decays are described by
\begin{equation}\label{eq::Leffcharged}
\mathcal L_{\rm eff}^{b\to c\tau\nu} = -\frac{4G_F}{\sqrt{2}}V_{cb}\sum_j^5 C_j\mathcal O_j\,,
\end{equation}
where the corresponding operators are defined as follows:
\begin{align}
\mathcal O_{V_{L,R}}&= (\bar c \gamma^\mu P_{L,R}b)\bar\tau \gamma_\mu \nu\,,&\mathcal O_{S_{L,R}}&=
(\bar c P_{L,R} b)\bar\tau\nu\,,\nonumber\\
\mathcal O_T &= (\bar c\sigma^{\mu\nu}P_L b)\bar\tau\sigma_{\mu\nu}\nu\,.&&
\end{align}
The structure of the occurring operators is constrained by the strong and electromagnetic symmetries.
Information about the full electroweak symmetry can be implemented by matching the previous set of
operators to effective field theories (EFTs) valid at the electroweak scale (see Refs.~\cite{D'Ambrosio:2002ex,Cirigliano:2009wk,Alonso:2014csa} for
earlier work in this direction). However, the specific form of the EFT at the electroweak scale strongly depends on the nature of the Higgs boson. For
a conventional SM Higgs, the scalar sector is described by a linear sigma model, where the Higgs and
the electroweak Goldstone bosons belong to a weak doublet. In order to {\it{test}} the SM
Higgs hypothesis, one needs a more general framework where the Goldstone bosons can be decorrelated from the Higgs particle. Such a framework is
provided by a nonlinear representation, see \emph{e.g.} Ref.~\cite{Feruglio:1992wf} for a general discussion. In its minimal implementation, the
Higgs is a scalar singlet and the electroweak Goldstone modes are contained in a matrix $U$ transforming as a bifundamental of $SU(2)_L\times SU(2)_R$, $U\to g_L U g_R^{\dagger}$. The linear and nonlinear representations
correspond to markedly different dynamical pictures of the mechanism triggering EWSB. The linear
EFT is suitable to describe weakly-coupled extensions of the SM, yielding an expansion in canonical
dimensions. The nonlinear EFT is in contrast aimed at strongly-coupled dynamics, and requires a loop expansion
(or equivalently an expansion in chiral dimensions~\cite{Buchalla:2013eza}).

Matching the nonlinear EFT to the flavor basis will therefore serve a two-fold purpose: (i)
it will ensure a model-independent implementation of the electroweak symmetry for flavor processes and (ii)
it will provide a framework to characterize departures from an SM Higgs. 

\vskip 0.2cm
\noindent{\it{2. Matching at the electroweak scale.}} 
We match the NLO nonlinear operator basis at the electroweak scale~\cite{Buchalla:2013rka} onto the
flavor EFT at $\Lambda\ll M_W$, providing at the same time the expressions for the (more restrictive)
linear basis~\cite{Buchmuller:1985jz,Grzadkowski:2010es}. The subset of NLO operators relevant for semileptonic processes is listed in the Appendix.
Hatted operators denote genuine nonlinear operators, while unhatted ones have counterparts at the same order in the linear basis.

We will start our discussion with the FCNC processes. For the dipole operators, the matching yields
\begin{align}
\delta C_{7(d)}^{(\prime)}&=\frac{8\pi^2}{m_b\lambda_{ts}}\frac{v^2}{\Lambda^2}\left[c_{X2}^{(\prime)}+c_{X4}^{(\prime)}\right]\,,\nonumber\\
\delta C_{7(u)}^{(\prime)}&=\frac{8\pi^2}{m_c\lambda_{bu}}\frac{v^2}{\Lambda^2}\left[c_{X1}^{(\prime)}+c_{X3}^{(\prime)}\right]\,,
\end{align}
while for the vector sector one finds (defining $\Nd=\frac{4\pi^2}{e^2\lambda_{ts}}\frac{v^2}{\Lambda^2}$ and
$\Nu=\frac{4\pi^2}{e^2\lambda_{bu}}\frac{v^2}{\Lambda^2}$)
\begin{align}
\delta
C_{9,10}^{(q)}&=\Nq\left[(C_{LR}^{(q)}\pm
C_{LL}^{(q)})\pm 4g_{V,A}\frac{\Lambda^2}{v^2}C_{VL}^{(q)}\right]\,,\nonumber\\
C_{9,10}^{\prime(q)}&=\Nq\left[(C_{RR}^{(q)}\pm
C_{RL}^{(q)})\pm 4g_{V,A}\frac{\Lambda^2}{v^2}C_{VR}^{(q)}\right]\,.
\end{align}
The first two contributions above correspond to local four-fermion contributions, whereas the last
one corresponds to $Z$-mediated diagrams, which become local at the heavy-flavor
scale.\footnote{Power-counting arguments show~\cite{Buchalla:2013rka} that $C_{VL,VR}^{(q)}$ carry a
one-loop suppression, while the remaining coefficients are ${\cal{O}}(1)$. Both contributions are
therefore of the same order.} The coefficients for the down- and up-quark sectors read 
\begin{align}
C_{LL}^{(d)}&=c_{LL1}+c_{LL2}-{\hat{c}}_{LL3}-{\hat{c}}_{LL4}+{\hat{c}}_{LL5}+{\hat{c}}_{LL6}-{\hat{c}}_{LL7}\,,\nonumber\\
C_{RR}^{(d)}&=c_{RR2}\,,~C_{LR}^{(d)}=c_{LR1}-{\hat{c}}_{LR5}\,,~C_{RL}^{(d)}=c_{LR3}-{\hat{c}}_{LR7}\,,\nonumber\\
C_{VL}^{(d)}&=c_{V1}-c_{V2}\,,~C_{VR}^{(d)}=c_{V4}\,,\quad\mbox{and}\\[1.3ex]
C_{LL}^{(u)}&=c_{LL1}-c_{LL2}+{\hat{c}}_{LL3}-{\hat{c}}_{LL4}-{\hat{c}}_{LL5}\,,\nonumber\\
C_{RR}^{(u)}&=c_{RR1}\,,~C_{LR}^{(u)}=c_{LR1}+{\hat{c}}_{LR5}\,,~C_{RL}^{(u)}=c_{LR2}-{\hat{c}}_{LR6}\,,\nonumber\\
C_{VL}^{(u)}&=c_{V1}+c_{V2}\,,~C_{VR}^{(u)}=c_{V3}\,.
\end{align}
The main conclusion to be drawn, independently of the EWSB mechanism, is that for the dipole
operators and the vector sector invariance under the electroweak symmetry does not add information compared to only imposing electromagnetic and
strong invariance, {\it{i.e.}}, the number of independent operators does not get reduced.  
This is not unexpected, since there are strong indications that these sectors are effectively decoupled from the mechanism of EWSB. In fact, in
the nonlinear framework they only appear as finite counterterms, {\it{i.e.}}, they are not needed to renormalize the EFT~\cite{Buchalla:2013rka}. As a
result they are expected to be rather insensitive to electroweak physics in general and to the existence of the Higgs in particular. 

The situation is substantially different for the scalar and tensor sectors. 
For the down-quark sector one
finds\footnote{Here and in the following, primed coefficients are related to unprimed coefficients
by taking the same sample operator listed in the Appendix, reverting the quark flavor indices and
applying hermitean conjugation.}
\begin{align}
C_{S,P}^{(d)}&=\Nd\left[\pm
c_{S}^{(d)}+\hat{c}_{Y1}\right]\,,&
C_{S,P}^{\prime (d)}&=\Nd\left[c^{\prime(d)}_{S}\pm\hat{c}_{Y1}^\prime\right]\,,\nonumber\\
C_T^{(d)}&=\Nd\left[\hat{c}_{Y2}+\hat{c}_{Y2}'\right]\,,&
C_{T5}^{(d)}&=\Nd\left[\hat{c}_{Y2}-\hat{c}_{Y2}'\right]\,,
\end{align}
where $c_{S}^{(\prime)(d)}=2(\hat{c}_{LR8}^{(\prime)}-c_{LR4}^{(\prime)})$. Notice that in the
linear case, {\it{i.e.}}, assuming a standard Higgs, one finds
\begin{align}\label{rel0}
C_S^{(d)}&=-C_P^{(d)}\,,&C^{\prime(d)}_S&=C^{\prime(d)}_P\,,&C^{(d)}_T&=C^{(d)}_{T5}=0\,,
\end{align}
as already noted in~\cite{Alonso:2014csa}. Deviations from these relations are expected to
arise at the percent level, through Higgs-exchange diagrams (which are NLO but numerically
suppressed by a small Yukawa coupling\footnote{If NP lies at the TeV scale, there is no generic
reason to expect it to be coupled to the SM fermions with Yukawa-like patterns. The local
contributions therefore dominate over the Higgs-exchange diagrams in general.}) and also by ${\cal{O}}(v^4/\Lambda^4)$ effects induced by NNLO operators,
all of which can be safely neglected. However, it is important to note that they are not predictions
of electroweak symmetry alone. For a nonstandard Higgs, one finds that the pattern of correlations
is already broken at NLO: the scalar sector is fully decorrelated and $C_T$ and $C_{T5}$ no longer
vanish. These decorrelations are caused by the operators ${\hat{\cal{O}}}_{Yj}$ which are,
contrary to the remaining operators listed in the Appendix, characterized by having
a fermion content with a nonvanishing total hypercharge; this is of course compensated by the
hypercharge of the $U$ field. For a standard Higgs such operators can only appear at NNLO and therefore provide a 
clean way of fingerprinting the nature of the Higgs boson.

A similar situation is encountered in the up-quark sector, where we obtain
\begin{align}
C_{S,P}^{(u)}&=\Nu\left[c_{S}^{(u)}\pm\hat{c}_{Y3}'\right]\,,&\!\!\!
C_{S,P}^{\prime(u)}&=\Nu\left[\pm c_{S}^{\prime(u)}+\hat{c}_{Y3}\right],\nonumber\\
C_T^{(u)}&=\Nu\left[c_{T}^{(u)}+c_{T}^{\prime(u)}\right]\,,&\!\!\!
C_{T5}^{(u)}&=\Nu\left[c_{T}^{(u)}-c_{T}^{\prime(u)}\right],
\end{align}
with $c_{S}^{(\prime)(u)}=-c_{S1}^{(\prime)}+\hat{c}_{S3}^{(\prime)}$ and
$c_T^{(\prime)(u)}=-c_{S2}^{(\prime)}+\hat{c}_{S4}^{(\prime)}$. A standard Higgs would in this case
predict
\begin{align}\label{corrU}
C_S^{(u)}&=C_P^{(u)}\,,&C_S^{\prime(u)}&=-C_P^{\prime(u)}\,.
\end{align}
Instead, for a nonstandard Higgs, Eqs.~(\ref{corrU}) are no longer satisfied due to the presence of
${\hat{c}}_{Y3}$. Due to the hypercharge structure of the transition, contributions to the tensor
sector appear already at NLO in both the linear and nonlinear EFTs and cannot be used to
discriminate between the two.

We now turn our attention to charged-current processes. For $b \to c\ell\nu$ transitions the matching
between the flavor and electroweak EFTs reads ($\Nbc=\frac{1}{2V_{cb}}\frac{v^2}{\Lambda^2}$)
\begin{align}
C_{V_L}&=-\Nbc\left[C_L+\frac{2}{v^2}c_{V5}+\frac{2V_{cb}}{v^2}c_{V7}\right]\,,\nonumber\\
C_{V_R}&=-\Nbc\left[\hat C_R+\frac{2}{v^2}c_{V6}\right]\,,\nonumber\\
C_{S_L}&=-\Nbc\,(c'_{S1}+\hat{c}'_{S5})\,,\nonumber\\
C_{S_R}&=2\,\Nbc\,(c_{LR4}+\hat{c}_{LR8})\,,\nonumber\\
C_T&=-\Nbc\,(c'_{S2}+\hat{c}'_{S6})\,,
\end{align}
where $C_L=2c_{LL2}-\hat{c}_{LL6}+\hat{c}_{LL7}$ and $\hat C_R=-\frac{1}{2}\hat{c}_{Y4}$. To simplify the notation, flavor rotation matrices have
been absorbed into the NLO coefficients of the EFT. This is always possible when considering a single transition.
However, when relating different processes, relative flavor rotations have to be taken into account explicitly, see below.

While in this case for both electroweak EFTs the full basis is reproduced with independent coefficients, correlations between {\emph{different}}
processes appear in the linear case with interesting consequences: (i) the absence of a direct four-fermion operator contribution to $C_{V_R}$ implies
lepton-flavor universality in that sector (inherited from the $W$-to-fermion couplings in the
SM), as already noted in~\cite{Cirigliano:2009wk}; (ii) the scalar sectors of the charged and neutral processes are related, see
{\it{e.g.}}~\cite{Cirigliano:2012ab}.
For instance,
\begin{align}\label{rel1}
\sum_{U=u,c,t}\lambda_{Us}C_{S_R}^{(U)}=-\frac{e^2}{8\pi^2}\lambda_{ts}C_S^{(d)}
\end{align} 
between $b\to U\ell\nu$ and $b\to s\ell\ell$, where we have
taken into account the relative flavor rotation to the quark mass eigenstates. Similar relations hold for all processes related by $SU(2)_L$.
We remark that this is non-trivial for the coefficients in the flavor EFT which generally receive contributions from several
electroweak invariant operators.
Incidentally, we note that Eq.~(\ref{rel1}) implies that bounds from FCNC processes severely constrain the size of new physics in their
$SU(2)$-related charged-current processes. 

We stress again that the correlations (i) and (ii) appear exclusively in the standard Higgs scenario and therefore can be used to test the
nature of the Higgs boson from flavor physics.
  
\vskip 0.2cm
\noindent{\it{3. Nontrivial hypercharge operators from TeV physics.}} The operators that most
prominently distinguish weakly- from strongly-coupled Higgs scenarios are
${\cal{\hat O}}_{Yj}$, {\it{i.e.}} four-fermion structures with nonvanishing fermionic hypercharge. In a
weakly-coupled scenario such structures can only appear at NNLO, {\it{e.g.}} $({\bar{q}}\varphi
d)(\bar{l}\varphi e)$ or $(\bar{l}\varphi d)({\bar{u}}{\tilde{\varphi}}^{\dagger}l)$, while in the
nonlinear case they are present already at NLO. Such operators can be generated in
simple models of heavy scalar exchanges, as we shall illustrate here.

As an example we consider an extension of the SM where EWSB is driven by a strongly-coupled
sector with the addition of two TeV-scale states: a scalar $\phi$, singlet under the SM group,
and a colored scalar $\Phi$, transforming as $({\bf{{\bar{3}}}},{\bf{2}})_{{\bf{-2/3}}}$. These
heavy states could be fundamental fields or composite states of the strong sector. Since we are
interested in four-fermion operators, we will concentrate on their couplings to fermionic scalar
currents:
\begin{align}\label{Lag}
{\cal{L}}_{int}(\phi,\Phi)&=\lambda_u{\bar{q}}UP_+\phi r+\lambda_d{\bar{q}}UP_-\phi r+\lambda_e{\bar{l}}UP_-\phi\eta\nonumber\\
&\!\!\!\!\!\!\!\!\!\!\!\!\!\!\!\!\!\!\!\!\!\!\!\!+\lambda_1 {\bar{l}}UP_+\Phi r+\lambda_2 {\bar{l}}UP_-\Phi r+\lambda_3 {\bar{\eta}}P_-U^{\dagger}\Phi q+{\mathrm{h.c.}}
\end{align}
For simplicity the coefficients are assumed to be real. Integrating out the heavy scalars,
the resulting EFT includes the terms 
\begin{align}
{\cal{L}}_{eff}&~\supset~\frac{\lambda_u\lambda_e}{2m_{\phi}^2}{\hat{\cal{O}}}_{Y3}+\frac{\lambda_u\lambda_e}{2m_{\phi}^2}{\hat{\cal{O}}}_{S3}+\frac{\lambda_d\lambda_e}{2m_{\phi}^2}{\hat{\cal{O}}}_{Y1}\nonumber\\
&-\frac{\lambda_d\lambda_e}{8m_{\phi}^2}\left({\cal{O}}_{LR4}-{\hat{\cal{O}}}_{LR8}\right)-\frac{\lambda_1^2}{8m_{\Phi}^2}({\cal{O}}_{LR2}+{\hat{\cal{O}}}_{LR6})\nonumber\\
&-\frac{\lambda_2^2}{8m_{\Phi}^2}({\cal{O}}_{LR3}-{\hat{\cal{O}}}_{LR7})-\frac{\lambda_3^2}{8m_{\Phi}^2}({\cal{O}}_{LR1}-{\hat{\cal{O}}}_{LR5})\nonumber\\
&+\frac{\lambda_1\lambda_2}{2m_{\Phi}^2}{\hat{\cal{O}}}_{Y4}-\frac{\lambda_1\lambda_3}{4m_{\Phi}^2}\left({\hat{\cal{O}}}_{S3}-\frac{1}{4}{\hat{\cal{O}}}_{S4}\right)\nonumber\\
&-\frac{\lambda_2\lambda_3}{4m_{\Phi}^2}\left({\hat{\cal{O}}}_{Y1}-\frac{1}{4}{\hat{\cal{O}}}_{Y2}\right)+{\mathrm{h.c.}}\,,
\end{align}
therefore generating explicitly the operators that violate the relations~\eqref{rel0}, \eqref{corrU} and \eqref{rel1}.

\vskip 0.2cm
\noindent{\it{4. Conclusions.}} We have analyzed the impact of the full electroweak symmetry at
hadronic scales, extending the work done in Refs.~\cite{Cirigliano:2009wk,Alonso:2014csa} and including
nonstandard Higgs scenarios. We have shown that correlations between different coefficients are
linked to assuming an SM ({\it{i.e.}}, weak doublet) Higgs field. Their violation would
point at a nonstandard Higgs, presumably of a composite nature. Flavor processes therefore
provide valuable information on the dynamics responsible for EWSB.
Furthermore, we have explored the consequences of assuming a linearly realized electroweak symmetry for the Higgs sector. This leads to
relations among the coefficients of the different flavor EFTs, thereby simplifying the corresponding model-independent analyses. 
Specifically, in addition to Eqs.~\eqref{rel0} for $D\to D'\ell\ell$ decays, already reported in Ref.~\cite{Alonso:2014csa}, we have found
Eqs.~\eqref{corrU} for $U\to U'\ell\ell$ decays. In charged-current decays, the electroweak symmetry implies lepton-flavor
universality for the right-handed vector currents. Furthermore, we pointed out the phenomenological significance of $SU(2)$ relations between FCNC and
charged-current semileptonic decays, {\it{e.g.}} Eq.~\eqref{rel1}. 

\vskip 0.2cm
\noindent{\it{Acknowledgments}}. We thank J.~M.~Camalich and M.~Gonzalez-Alonso for useful comments on the manuscript.
This work was performed in the context of the ERC Advanced Grant project 'FLAVOUR' (267104) and was supported in part by the DFG cluster of excellence
'Origin and Structure of the Universe'.

\vskip 0.2cm 
\noindent{\it{Appendix.}} In the following we list the subset of relevant operators for semileptonic
processes, extracted from the NLO operator basis worked out in Ref.~\cite{Buchalla:2013rka}.
Notational changes have been introduced to ease the comparison with the linear basis in
Ref.~\cite{Grzadkowski:2010es}. 
For simplicity, we will work in unitary gauge, where $U=1$. Below we will however keep $U$ in order to make the invariance under the SM gauge symmetry transparent. $\hat{\tau}_3=U\tau_3U^{\dagger}$ and ${\hat{\tau}}_{\pm}=U\frac{1}{2}(\tau_1\pm
i\tau_2)U^{\dagger}$ are chirally-dressed Pauli matrices, and $L_\mu\equiv iUD_\mu U^\dagger$.

The subset of operators relevant for the electromagnetic dipole operators are
\begin{align}
{\cal{O}}_{X1,2}&=g'{\bar{q}}\sigma^{\mu\nu}UP_\pm rB_{\mu\nu},&\!\!\!\!
{\cal{O}}_{X3,4}&=g{\bar{q}}\sigma^{\mu\nu}UP_\pm r\langle \hat{\tau}_3
W_{\mu\nu}\rangle,\nonumber\\
{\cal{O}}_{X1,2}'&=g'{\bar{r}}P_\pm U^{\dagger}\sigma^{\mu\nu}qB_{\mu\nu},&\!\!\!\!
{\cal{O}}_{X3,4}'&=g{\bar{r}}P_\pm U^{\dagger}\sigma^{\mu\nu}q\langle \hat{\tau}_3
W_{\mu\nu}\rangle,\nonumber
\end{align} 
while the one for the vector operators reads
\begin{align}
{\cal{O}}_{V1}&={\bar{q}}\gamma^{\mu}q\langle \hat{\tau}_3 L_{\mu}\rangle\,,&
{\cal{O}}_{V2}&={\bar{q}}\gamma^{\mu}\hat{\tau}_3 q\langle \hat{\tau}_3 L_{\mu}\rangle\,,\nonumber\\
{\cal{O}}_{V3}&={\bar{u}}\gamma^{\mu}u\langle\hat{\tau}_3 L_{\mu}\rangle\,,&
{\cal{O}}_{V4}&={\bar{d}}\gamma^{\mu}d\langle \hat{\tau}_3 L_{\mu}\rangle\,,\nonumber\\
{\cal{O}}_{V5}&={\bar{q}}\gamma^{\mu}\hat{\tau}_+ q\langle \hat{\tau}_- L_{\mu}\rangle\,,&
{\cal{O}}_{V6}&={\bar{u}}\gamma^{\mu}d\langle \hat{\tau}_- L_{\mu}\rangle\,,\nonumber\\
{\cal{O}}_{V7}&={\bar{l}}\gamma^{\mu}\hat{\tau}_- l\langle \hat{\tau}_+
L_{\mu}\rangle\,,\nonumber\\[1.ex]
{\cal{O}}_{LL1}&={\bar{q}}\gamma^{\mu}q~{\bar{l}}\gamma_{\mu}l\,,& {\cal{O}}_{LL2}&={\bar{q}}\gamma^{\mu}\tau^jq~{\bar{l}}\gamma_{\mu}\tau^jl\,,\nonumber\\
{\cal{\hat{O}}}_{LL3}&={\bar{q}}\gamma^{\mu}\hat{\tau}_3q~{\bar{l}}\gamma_{\mu}l\,,&
{\cal{\hat{O}}}_{LL4}&={\bar{q}}\gamma^{\mu}q~{\bar{l}}\gamma_{\mu}\hat{\tau}_3l\,,\nonumber\\
{\cal{\hat{O}}}_{LL5}&={\bar{q}}\gamma^{\mu}\hat{\tau}_3q~{\bar{l}}\gamma_{\mu}\hat{\tau}_3l\,,&
{\cal{\hat{O}}}_{LL6}&={\bar{q}}\gamma^{\mu}\hat{\tau}_3l~{\bar{l}}\gamma_{\mu}\hat{\tau}_3q\,,\nonumber\\
{\cal{\hat{O}}}_{LL7}&={\bar{q}}\gamma^{\mu}\hat{\tau}_3l~{\bar{l}}\gamma_{\mu}q\nonumber\,,\\[1.3ex]
{\cal{O}}_{LR1}&={\bar{q}}\gamma^{\mu}q~{\bar{e}}\gamma_{\mu}e\,,&
{\cal{O}}_{LR2}&={\bar{u}}\gamma^{\mu}u~{\bar{l}}\gamma_{\mu}l\,,\nonumber\\
{\cal{O}}_{LR3}&={\bar{d}}\gamma^{\mu}d~{\bar{l}}\gamma_{\mu}l\,,&
{\cal{\hat{O}}}_{LR5}&={\bar{q}}\gamma^{\mu}\hat{\tau}_3q~{\bar{e}}\gamma_{\mu}e\,,\nonumber\\
{\cal{\hat{O}}}_{LR6}&={\bar{u}}\gamma^{\mu}u~{\bar{l}}\gamma_{\mu}\hat{\tau}_3l\,,&
{\cal{\hat{O}}}_{LR7}&={\bar{d}}\gamma^{\mu}d~{\bar{l}}\gamma_{\mu}\hat{\tau}_3l\,,\nonumber\\[1.3ex]
{\cal{O}}_{RR1}&={\bar{u}}\gamma^{\mu}u{\bar{e}}\gamma_{\mu}e\,,&
{\cal{O}}_{RR2}&={\bar{d}}\gamma^{\mu}d~{\bar{e}}\gamma_{\mu}e\,.\nonumber
\end{align}

Finally, operators relevant for the scalar and tensor sectors at the electroweak scale are
\begin{align}
{\cal{O}}_{LR4}&={\bar{q}}\gamma^{\mu}l~{\bar{e}}\gamma_{\mu}d\,,\!\!\!&{\cal{\hat{O}}}_{LR8}&={\bar{q}}\gamma^{\mu}\hat{\tau}_3l~{\bar{e}}\gamma_{\mu}d\,,\nonumber\\[1.3ex]
{\cal{O}}_{S1}&=\epsilon_{ij}{\bar{l}}^ie{\bar{q}}^ju\,,\!\!\!&
{\cal{O}}_{S2}&=\epsilon_{ij}{\bar{l}}^i\sigma^{\mu\nu}e{\bar{q}}^j\sigma_{\mu\nu}u\,,\nonumber\\
{\cal{\hat{O}}}_{S3}&={\bar{q}}UP_+r{\bar{l}}UP_-\eta\,,\!\!\!&
{\cal{\hat{O}}}_{S4}&={\bar{q}}\sigma_{\mu\nu}UP_+r{\bar{l}}\sigma^{\mu\nu}UP_-\eta\,,\nonumber\\
{\cal{\hat{O}}}_{S5}&={\bar{q}}\hat{\tau}_-Ur{\bar{l}}\hat{\tau}_+U\eta\,,\!\!\!&
{\cal{\hat{O}}}_{S6}&={\bar{q}}\sigma_{\mu\nu}\hat{\tau}_-Ur{\bar{l}}\sigma^{\mu\nu}\hat{\tau}_+U\eta\,,\nonumber\\[1.3ex]
{\cal{\hat{O}}}_{Y1}&={\bar{q}}UP_-r{\bar{l}}UP_-\eta\,,\!\!&
{\cal{\hat{O}}}_{Y2}&={\bar{q}}\sigma_{\mu\nu}UP_-r{\bar{l}}\sigma^{\mu\nu}UP_-\eta\,,\nonumber\\
{\cal{\hat{O}}}_{Y3}&={\bar{l}}UP_-\eta{\bar{r}}P_+U^{\dagger}q\,,\!\!&
{\cal{\hat{O}}}_{Y4}&={\bar{l}}UP_-r{\bar{r}}P_+U^{\dagger}l\,.\nonumber
\end{align}
Flavor family indices have been omitted.


\end{document}